\documentstyle[12pt,psfig,draft,lscape,array,amssymb]{nature-ejb-new}

\def\simlt{\mathrel{\hbox{\rlap{\hbox{\lower4pt\hbox{$\sim$}}}\hbox{$<$}}}}
\def\simgt{\mathrel{\hbox{\rlap{\hbox{\lower4pt\hbox{$\sim$}}}\hbox{$>$}}}}

\def\ale{\mathrel{\hbox{\rlap{\hbox{\lower4pt\hbox{$\sim$}}}\hbox{$<$}}}}
\def\age{\mathrel{\hbox{\rlap{\hbox{\lower4pt\hbox{$\sim$}}}\hbox{$>$}}}}

\def\nodata{---}

\def\ra#1#2#3{#1$^{\rm h}$#2$^{\rm m}$#3$^{\rm s}$}
\def\dec#1#2#3{$#1^\circ#2'#3''$}

\newcommand{\swift}{\textit{Swift}}

\def\spose#1{\hbox to 0pt{#1\hss}}
\newcommand\lsim{\mathrel{\spose{\lower 3pt\hbox{$\mathchar"218$}}
     \raise 2.0pt\hbox{$\mathchar"13C$}}}
\newcommand\gsim{\mathrel{\spose{\lower 3pt\hbox{$\mathchar"218$}}
     \raise 2.0pt\hbox{$\mathchar"13E$}}}

\begin{document}

\title{\Large \bf Relativistic ejecta from XRF\,060218 and the rate of cosmic explosions}

\author{A.~M.~Soderberg\affiliation[1]
  {Caltech Optical Observatories 105-24, California Institute of
       Technology, Pasadena, CA 91125, USA},
   S.~R.~Kulkarni\affiliationmark[1],
   E.~Nakar\affiliation[2]
   {Theoretical Astrophysics 130-33, California Institute of
       Technology, Pasadena, CA 91125, USA},
   E.~Berger\affiliation[3]
   {Carnegie Observatories, 813 Santa Barbara St., Pasadena, CA
        91101, USA},
   P.~B.~Cameron\affiliationmark[1],
   D.~B.~Fox\affiliation[4]
   {Department of Astronomy, Pennsylvania State University, University Park, PA 16802, USA},
   D.~Frail\affiliation[5]
   {National Radio Astronomy Observatory, P.O. Box 0, Socorro, New
      Mexico 87801, USA},
   A.~Gal-Yam\affiliationmark[1],
   R.~Sari\affiliationmark[1],
   S.~B.~Cenko\affiliation[6]
   {Space Radiation Laboratory 220-47, California Institute of
      Technology, Pasadena, CA 91125, USA},
   M.~Kasliwal\affiliationmark[1],
   R.~A.~Chevalier\affiliation[7]
     {Department of Astronomy, University of Virginia, PO Box 3818, Charlottesville, VA 22903, USA},
   T.~Piran\affiliation[8]
     {Racah Institute of Physics, Hebrew University, Jerusalem
      91904, Israel},
   P.~A.~Price\affiliation[9]
     {Institute for Astronomy, University of Hawaii, 2680 Woodlawn
      Drive, Honolulu, HI 96822, USA},
   B.~P.~Schmidt\affiliation[10]
     {RSAA, ANU, Mt.\ Stromlo Observatory, via Cotter Rd, Weston
      Creek, ACT 2611, Australia},
   G.~Pooley\affiliation[11]
     {Mullard Radio Astronomy Observatory, Cavendish Laboratory, Cambridge CB3 0HE, UK},
   D.-S.~Moon\affiliationmark[6],
   B.~E.~Penprase\affiliation[12]
     {Pomona College Dept.~of Physics \& Astronomy, 610 N.~College Ave,
      Claremont, CA 91711, USA},
   E.~Ofek\affiliationmark[1],
   A.~Rau\affiliationmark[1],
   N.~Gehrels\affiliation[13]
     {NASA Goddard Space Flight Center, Greenbelt, MD 20771, USA},
   J.~A.~Nousek\affiliationmark[4],
   D.~N.~Burrows\affiliationmark[4],
   S.~E.~Persson\affiliationmark[3], and
   P.~J.~McCarthy\affiliationmark[3]
}
\date{\today}{}
\headertitle{XRF\,060218}
\mainauthor{Soderberg et al.}

\summary{ Over the last decade, long-duration $\gamma$-ray bursts
  (GRBs) including the subclass of X-ray flashes (XRFs) have been
  revealed\cite{gvv+98,kfw+98,mgs+03} to be a rare variety of Type Ibc
  supernova (SN).  While all these events result from the death of
  massive stars, the electromagnetic luminosities of GRBs and XRFs
  exceed those of ordinary Type Ibc SNe by many orders of magnitude.
  The essential physical process that causes a dying star to produce a
  GRB or XRF, and not just an SN, remains the crucial open question.
  Here we present radio and X-ray observations of XRF\,060218
  (associated\cite{pmm+06} with SN\,2006aj), the second
  nearest\cite{cmb+06,mha+06} GRB identified to-date, which allow us
  to measure its total energy and place it in the larger context of
  cosmic explosions.  We show that this event is 100 times less
  energetic but ten times more common than cosmological GRBs.
  Moreover, it is distinguished from ordinary Type Ibc SNe by the
  presence of $10^{48}$ erg coupled to mildly-relativistic ejecta,
  along with a central engine (an accretion-fed, rapidly rotating
  compact source) which produces X-rays for weeks after the explosion.
  This suggests that the production of relativistic ejecta is the key
  physical distinction between GRBs/XRFs and ordinary SNe, while the
  nature of the central engine (black hole or magnetar) may
  distinguish typical bursts from low-luminosity, spherical
  events like XRF\,060218. }

\maketitle


On 2006 February 18.15 UT, the Burst Alert Telescope (BAT), a hard
X-ray detector aboard the \swift\ satellite, detected\cite{cmb+06} an
exceedingly long-duration ($\Delta t\approx 2000$ sec) transient.
Within 153 seconds of the $\gamma$-ray trigger, the on-board X-ray
Telescope (XRT) and Ultra-Violet Optical Telescope (UVOT)
identified\cite{cmb+06} a counterpart coincident\cite{mha+06} with a
dwarf galaxy at $z=0.0335$.  The XRT and BAT data show\cite{cmb+06}
that the event peaked at a photon energy of 4.9 keV, thereby
classifying this transient as an X-ray Flash, XRF\,060218.
Distinguished\cite{hik+01} by their soft X-ray dominated spectrum
(peak energy, $E_{p}\lesssim 25$ keV versus 250 keV), the
subclass of XRFs are otherwise similar (see ref.~\pcite{skb+04a} and
references therein) to GRBs in their observational properties.

Using the Very Large Array (VLA), we discovered a radio source at
$\alpha$(J2000)=\ra{03}{21}{39.68} and
$\delta$(J2000)=\dec{16}{52}{01.82} ($\pm 0.02$ arcsec in each axis),
coincident with the UVOT position.  Our monitoring of the radio source
showed a power-law decay with $\alpha\approx -0.8$ through $t\approx
22$ d (Table~\ref{tab:vla}), similar to the decay of afterglows seen
from GRBs; here $F_{\nu}\propto t^{\alpha}$ is the spectral flux
density.  Over the same period the XRT undertook intensive observations
of the source in the X-ray band (0.3--10\,keV). We find the X-ray
spectral flux density, $F_{\nu, X} \propto \nu^{\beta_X}$, is fit by
$\beta_X=-2.2\pm 0.2$ with an absorbing column density, $N_{\rm H} =
3.9\pm 0.4\times 10^{21}~\rm cm^{-2}$, consistent with previously
reported\cite{cmb+06,d06} values.

Separately, we observed the source with the {\it Advanced CCD Imaging
  Spectrometer} (ACIS) instrument aboard the {\it Chandra} X-ray
Observatory ({\it CXO}).  These observations began on 2006 February
26.78 and March 7.55 UT ($t\approx 8.8$ and 17.4 days) and lasted
about 20 and 30 ks, respectively.  The measured count rates are
$(1.9\pm 0.3)\times 10^{-3}$ and $(1.3\pm 0.3)\times
10^{-3}$\,s$^{-1}$, respectively.  Using the XRT model parameters
stated above we derive $F_X=(4.5\pm 1.4)\times 10^{-14}$ and $(2.8\pm
0.9)\times 10^{-14}~\rm erg~cm^{-2}~s^{-1}$ for the unabsorbed flux
values.  The XRT-{\it CXO} data spanning the range from a few minutes
to 17\,d are well fit a simple power-law decay model with temporal
index, $\alpha_X=-1.1$.

                        

XRF\,060218 is most interesting because it is nearby, distance
$d\approx 145\,$Mpc.  Indeed it is second only to
GRB\,980425/SN~1998bw\cite{gvv+98} at just 36\,Mpc.
Similar to GRB\,980425, XRF\,060218 is also associated\cite{pmm+06}
with a Type Ic supernova explosion, SN\,2006aj.  The isotropic
prompt energy release\cite{cmb+06} $E_{\gamma,\rm iso}=(6.2\pm
0.3)\times 10^{49}$ erg, is at least 100 times fainter than typical
GRBs but comparable to another nearby event,
GRB\,031203\cite{sls04,skb+04b} ($z=0.106$).  Similarly, the radio and
X-ray luminosities are $10^3$ and $10^2$ times fainter than those of
cosmological GRBs, respectively.

\begin{figure}
\centerline{\psfig{file=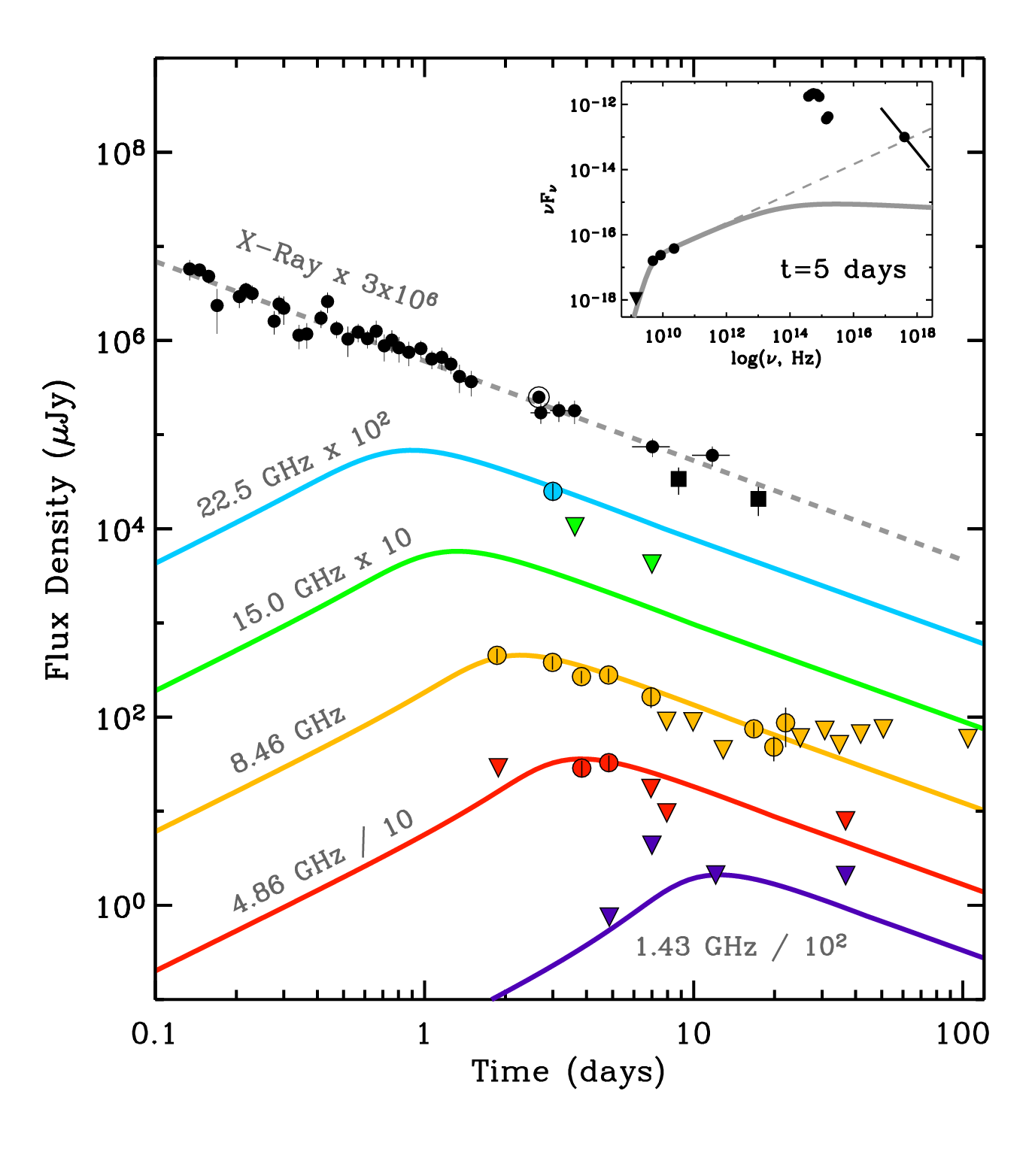,width=6.in,angle=0}}
\bigskip
\bigskip
\caption[]{}
\label{fig:oneplot2}
\end{figure}

Radio observations directly probe the ejecta and environments of
stellar explosions since the blastwave (velocity $v$) shocks the
circumstellar medium and accelerates relativistic electrons which give
rise to radio synchrotron emission.  For radio sources dominated by
synchrotron-self absorption, the brightness temperature is $T_B \simlt
4\times 10^{10}$ K. As can be seen from Figure~\ref{fig:oneplot2},
at day 5 the radio emission peaks between 1.4\,GHz and
4.9\,GHz. Applying the basic equipartition analysis (see
ref.~\pcite{kfw+98}) we find, at this epoch, that the radius of the radio
emitting region is $r\approx 3\times 10^{16}\,$cm, the ejecta kinetic
energy is $E_{K}\approx 2\times 10^{48}$ erg and the circumburst
particle density is $n\approx 5~\rm cm^{-3}$.  The blast wave thus
expands with a Lorentz factor $\Gamma=(1-\beta^2)^{-1/2}\sim 2.3$;
here $\beta\equiv v/c$.

\begin{figure}
\centerline{\psfig{file=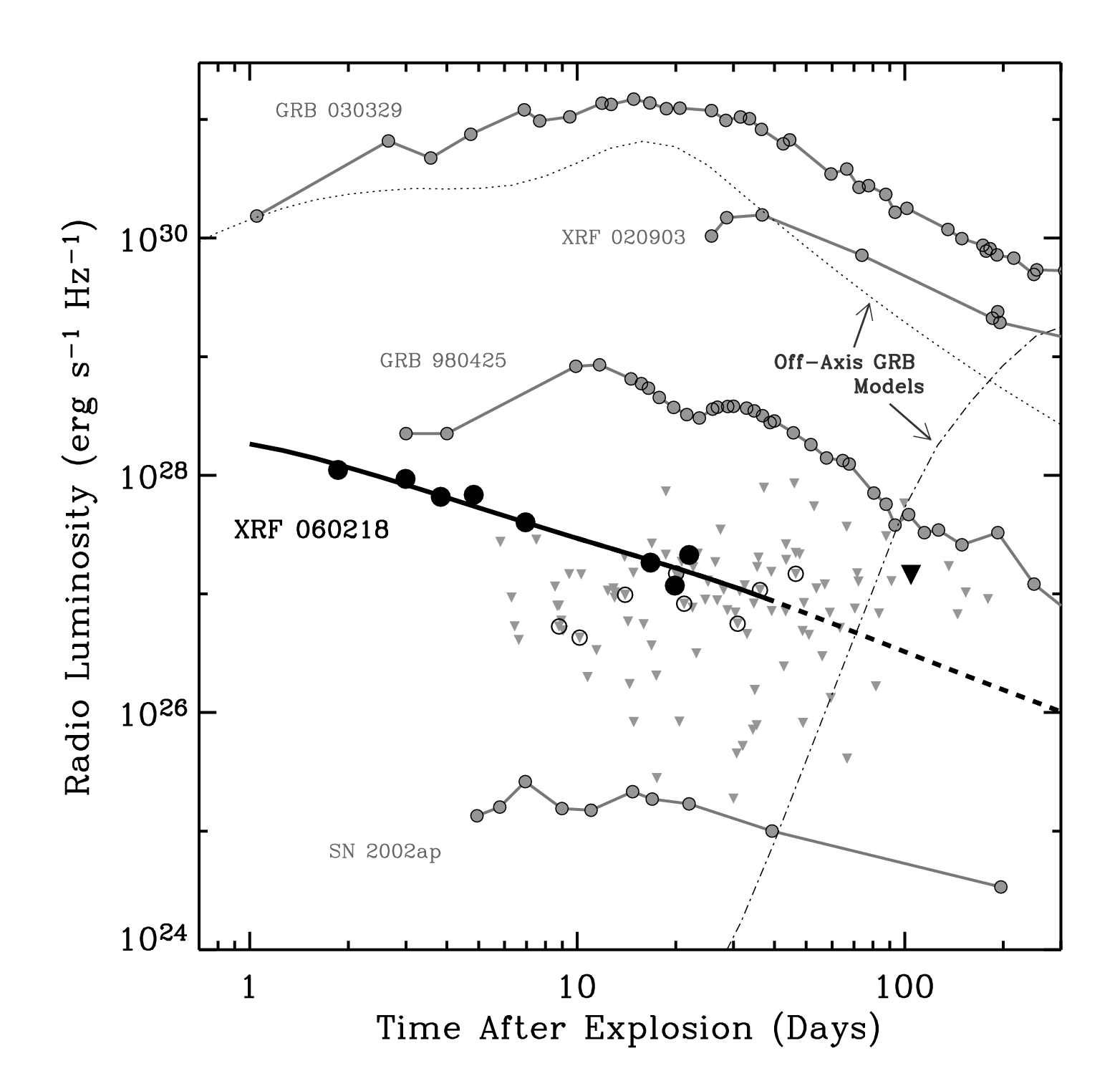,width=6.in,angle=0}}
\bigskip
\bigskip
\caption[]{}
\label{fig:lum_limits_060218}
\end{figure}

The early, steady decay of the radio emission indicates\cite{snb+06}
that it cannot be attributed to a collimated jet directed away from
our line-of-sight.  Moreover, on a timescale, $t_{\rm NR}\approx
7.3(E_{\rm K,48}/n_0)^{1/3}\,$days, the blastwave becomes\cite{wax04}
sub-relativistic ($\Gamma\beta<1$) at which point it effectively
assumes spherical geometry, even if the initial explosion was
biconical.  Independently, noting the absence of a ``jet break'' in
the radio light-curve (to 22\,d) and applying the standard
formulation\cite{sph99} we find the opening angle, $\theta_j\gtrsim
1.4$ radian. Thus, on several grounds, the radio data argue for a
quasi-spherical ejecta with $10^{48}$ erg coupled to
mildly-relativistic material.  In addition, our observations at 104
days show no evidence for a late-time increase in the radio flux, thus
constraining the presence of additional ejecta components (off-axis
jets; Figure~\ref{fig:lum_limits_060218}) spreading into our line-of-sight.


As can be seen from Figure~\ref{fig:oneplot2} the above synchrotron
model is unable to explain the strong X-ray emission.  Attributing the
emission to scattering of SN optical photons by the
mildly-relativistic ejecta requires an optical depth, $\tau\sim
10^{-4}$, too large to be produced by the shocked electrons which
provide $\tau=nr\sigma_T\sim 10^{-7}$; here $\sigma_T$ is the
Thomson cross section.  We must therefore seek an entirely different
origin for the observed X-rays.

At day 1, the steep X-ray spectrum roughly connects to the peculiar
optical/UV component ($\beta_{OX}\sim -2$) observed\cite{cmb+06} to
peak on this timescale.  A similar steep near-IR spectrum was
seen\cite{mtc+04} in GRB\,031203 at $t=0.4\,$d.  Given that both
GRB\,031203 and XRF\,060218 are\cite{skb+04b,sls04} sub-energetic
events, we suggest that this mysterious steep component is ubiquitous
among sub-energetic GRBs and speculate that a central engine is the
origin of this intense, long-lived emission. One particularly
attractive possibility is a rapidly rotating (period, $P$) highly
magnetized (field strength, $B$) neutron star, a {\em magnetar}. The
spin down power, $\dot E=10^{45}{\rm (P/10\,ms)^{-4}(B/10^{15}\,\rm
G)^2\,erg\,s^{-1}}$, can explain the peculiar optical to X-ray
integrated luminosity at 1 day while the temporal evolution requires a
braking index lower than three (magnetic dipole). We note that
similarly low braking indices are measured for young Galactic pulsars
(e.g. ref.~\pcite{lpg+96} and references therein).

Combining the sky coverage and detection thresholds of $\gamma-$ray
missions, we estimate the following sensitivity to the two exemplars
of low energy events (GRB\,980425 and XRF\,060218): $3.8\times
10^{-3}$ ({\em BeppoSAX}), $1.2\times 10^{-3}$ ({\it HETE-2}) and
$3.7\times 10^{-3}~\rm Gpc^{3}~yr$ ({\em Swift}).  Thus, the true rate
of sub-energetic GRBs is $230^{+490}_{-190}~\rm Gpc^{-3}~yr^{-1}$
($90\%$ confidence range; see Supplementary Information I), about 10
times more abundant than typical bright GRBs\cite{sch01}, for which we
use a mean inverse beaming factor of $<f_b^{-1}>\sim 100$; here
$f_b\equiv 1-{\rm cos}\theta_j$.  Separately, we note that
sub-energetic GRBs could not be strongly beamed or the true rate of
such events would exceed the local ($d\lesssim 100$ Mpc)
rate\cite{cet99,dsr+04} of Type Ibc supernovae, $9^{+3}_{-5}\times
10^3~\rm Gpc^{-3}~yr^{-1}$.

Spectroscopy of the nearest GRB-associated supernovae (SNe 1998bw,
2003dh, 2003lw and now 2006aj) reveals (see ref.~\pcite{pmm+06} and
references therein) remarkably broad absorption lines (indicative of
fast ejecta) and may suggest that all GRB-SNe are broad-lined (BL).
Locally, BL events comprise\cite{pmn+04} five percent of SNe Ibc.
Thus, the rate of BL events and sub-energetic GRBs are comparable,
suggesting that all BL SNe harbor a long-lived central engine.

Radio observations of an extensive sample of 144 optically-selected
local SNe Ibc, however, suggest\cite{bkf+03,snb+06} a different
picture (Figure~\ref{fig:lum_limits_060218}).  Not a single SN (BL or
otherwise) shows strong early radio emission comparable to that seen
in SNe 1998bw and 2006aj.  Thus, we constrain the volumetric rate of
such events to be $\lesssim 300~\rm Gpc^{-3}~yr^{-1}$ (see
Supplementary Information II), consistent with the rate of sub-energetic
GRBs inferred above.  Focusing on the BL SNe, less than one in three
are similar to GRBs, indicating that broad lines cannot be used as a
reliable proxy for a central engine.

\begin{figure}
\centerline{\psfig{file=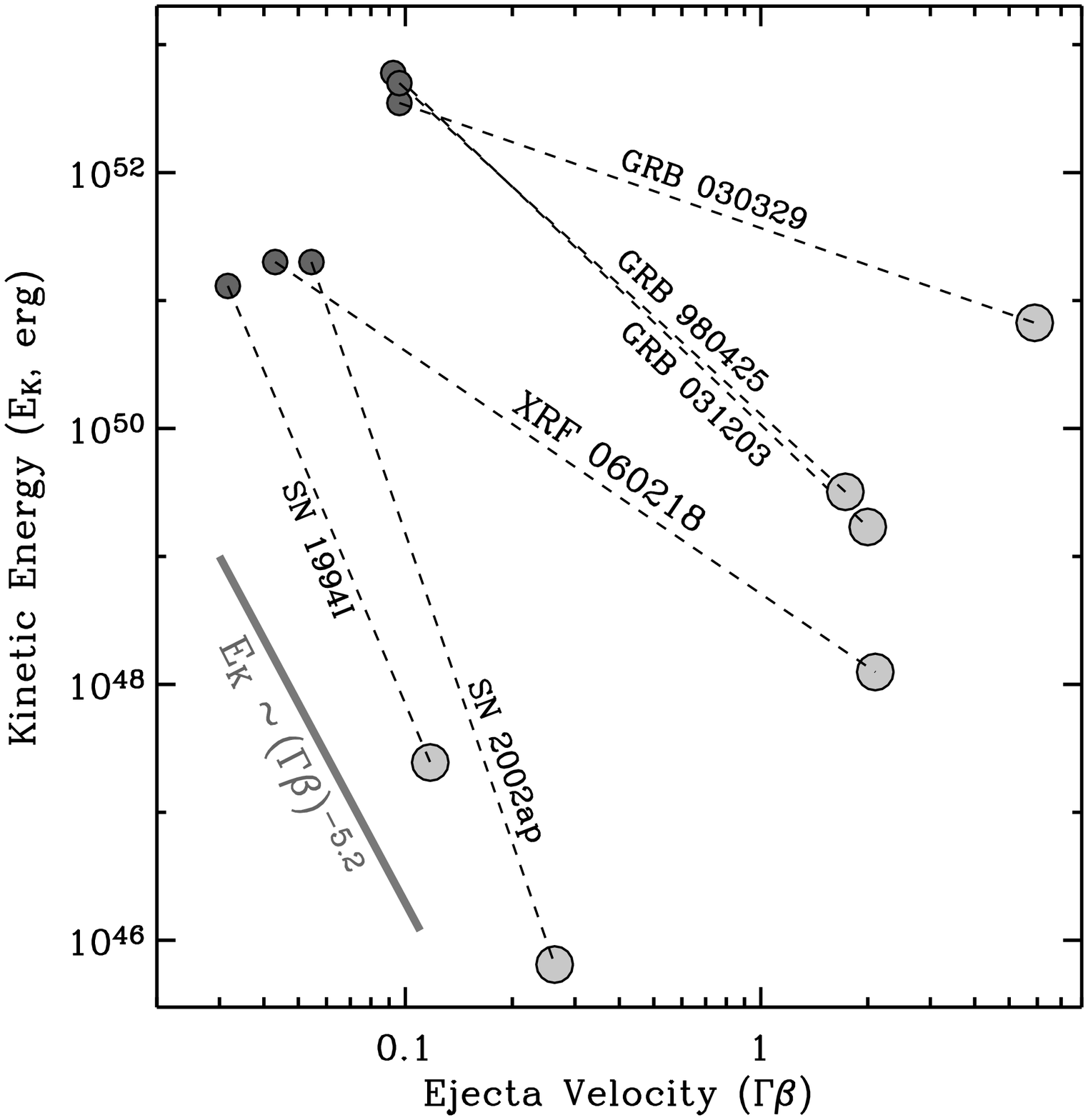,width=6in,angle=0}}
\bigskip
\bigskip
\caption[]{}
\label{fig:epep}
\end{figure}

The commonality between the three nearest events (980425, 060218,
031203) is their substantial ($E_K\simgt 10^{48}\,$erg)
mildly-relativistic ($\Gamma \gtrsim 2$) ejecta and a smooth pulse
profile for the prompt emission.  These two clues lead us to suggest
that the primary {\it physical} distinction between GRBs/XRFs and ordinary
supernovae is the velocity profile of the ejecta.  For the latter,
hydrodynamic collapse requires that the ejecta energy is concentrated
at low velocities, $E_K\propto (\Gamma\beta)^{-5.2}$.  In comparison,
the shallow velocity profiles inferred for GRBs and XRFs indicate that
some other agent (an engine) enables coupling of copious energy to
to relativistic material (Figure~\ref{fig:epep}).

We conclude by noting that magnetars constitute\cite{gmo+05} about
10\% of the Galactic neutron star birth-rate, and thus a similar
fraction of SNe Ibc.  This rate is similar to that of the
sub-energetic GRBs.  Furthermore, magnetars produce long-lived
emission (see ref.~\pcite{hbs+05} and references therein) and have
been suggested\cite{uso92} previously as candidate GRB progenitors.
We therefore speculate that a magnetar central engine is what
distinguishes sub-energetic GRBs from the cosmological bursts, which
are thought to be powered by a black hole.

\bibliographystyle{nature-pap}

\noindent
Correspondence should be addressed to A. M. Soderberg
(e-mail:ams@astro.caltech.edu).

\begin{acknowledge}
GRB research at Caltech is supported in part by funds from NSF and
NASA.  We are, as always, indebted to Scott Barthelmy and the GCN.
The VLA is operated by the National Radio Astronomy Observatory, a
facility of the National Science Foundation operated under cooperative
agreement by Associated Universities, Inc.  AMS and SBC are supported
by NASA Graduate Research Fellowships.  EB and AG acknowledge support
by NASA through a Hubble Fellowship grant.  DNB and JAN acknowledge
support by NASA.

\end{acknowledge}

\clearpage

\begin{table}
\begin{center}
\setlength{\extrarowheight}{-0.075in}
\begin{tabular}{>{\scriptsize}l >{\scriptsize}c >{\scriptsize}c >{\scriptsize}c >{\scriptsize}c >{\scriptsize}c >{\scriptsize}c}
\hline
\hline

Epoch & $\Delta t$ & $F_{1.43}$ & $F_{4.86}$ & $F_{8.46}$ &
$F_{15.0}$ & $F_{22.5}$ \\
(UT) & (days) & ($\mu$Jy) & ($\mu$Jy) & ($\mu$Jy) & ($\mu$Jy) & ($\mu$Jy) \\
\hline 
2006 Feb 20.02 & 1.87 & \nodata & $78\pm 70$ & $453\pm 77$ & \nodata & \nodata \\
2006 Feb 21.14 & 3.00 & \nodata & \nodata & $381\pm 60$ & \nodata & $250\pm 52$ \\ 
2006 Feb 21.77$^{\dagger}$ & 3.62 & \nodata & \nodata & \nodata & $350\pm 350$ & \nodata \\ 
2006 Feb 21.97 & 3.83 & \nodata & $287\pm 56$ & $269\pm 40$ & \nodata & \nodata \\
2006 Feb 22.99 & 4.85 & $25\pm 25$ & $328\pm 61$ & $280\pm 47$ & \nodata & \nodata \\
2006 Feb 25.12 & 6.97 & $134\pm 145$ & $80\pm 47$ & $164\pm 39$ & $46\pm 141$ & \nodata \\
2006 Feb 26.09 & 7.94 & \nodata & $32\pm 32$ & $30\pm 30$ & \nodata & \nodata \\
2006 Feb 28.10 & 9.95 & \nodata & \nodata & $39\pm 25$ & \nodata & \nodata \\ 
2006 Mar 2.23 & 12.08 & $70\pm 70$ & \nodata & \nodata & \nodata & \nodata \\
2006 Mar 3.03 & 12.88 & \nodata & \nodata & $15\pm 15$ & \nodata & \nodata \\
2006 Mar 6.89 & 16.74 & \nodata & \nodata & $75\pm 13$ & \nodata & \nodata \\ 
2006 Mar 10.01 & 19.86 & \nodata & \nodata & $48\pm 14$ & \nodata & \nodata \\
2006 Mar 12.11 & 21.96 & \nodata & \nodata & $87\pm 39$ & \nodata & \nodata \\
2006 Mar 15.04 & 24.91 & \nodata & \nodata & $20\pm 20$ & \nodata & \nodata \\ 
2006 Mar 20.86 & 30.71 & \nodata & \nodata & $32\pm 20$ & \nodata & \nodata \\
2006 Mar 24.96 & 34.81 & \nodata & \nodata & $15\pm 18$ & \nodata & \nodata \\
2006 Mar 26.85 & 36.70 & $69\pm 69$ & $5\pm 37$ & \nodata & \nodata & \nodata \\
2006 Mar 31.89 & 41.74 & \nodata & \nodata & $22\pm 22$ & \nodata & \nodata \\
2006 Apr 9.84 & 50.70 & \nodata & \nodata & $25\pm 25$ & \nodata & \nodata \\ 
2006 Jun 2.67 & 104.52 & \nodata & \nodata & $17\pm 21$ & \nodata & \nodata \\
\end{tabular}
\end{center}
\caption[]{}
\label{tab:vla}
\end{table}

\small Radio observations made with the Very Large Array
  (VLA) and the Ryle Telescope$^{\dagger}$.  We used the standard
  continuum mode with $2\times 50$ MHz bands (VLA) and 350 MHz
  bandwidth (Ryle).  At 22.5 GHz we used referenced pointing scans to
  correct for the systematic $10 - 20$ arcsec pointing errors of the
  VLA antennas.  We used the extra-galactic sources 3C\,48 (J0137+331)
  and 3C\,147 (J0542+498) for flux calibration, while the phase was
  monitored using J0319+190 (VLA) and J0326+1521 (Ryle).  The data
  were reduced and analyzed using the Astronomical Image Processing
  System.  The flux density and uncertainty were measured from the
  resulting maps by fitting a Gaussian model to the afterglow
  emission.

\clearpage

\noindent
{\bf Figure 1}: {\small Radio and X-ray light-curves of XRF\,060218.
  Radio measurements are summarized in Table~\ref{tab:vla}.  Upper
  limits are given as $3\sigma$ (inverted triangles). Solid lines are
  models of synchrotron emission from a spherical shock expanding into
  a wind-blown circumstellar medium ($n\propto r^{-2}$).  At $t=5$ d
  the radio spectrum peaks near 4 GHz due to the synchrotron
  self-absorption frequency, $\nu_a$.  We assume that the energy
  density is partitioned between the relativistic electrons (energy
  distribution $N(\gamma) \propto \gamma^{-p}$ with $p\approx 2.1$)
  and magnetic field as $\epsilon_e=\epsilon_B=0.1$.  We find that
  $E_K\approx 2\times 10^{48}$ erg is coupled to ejecta with
  $\Gamma\approx 2.3$.  The expansion, $r\propto t^m$,
  appropriate\cite{c98} for a core-collapse SN explosion with a
  distribution of ejecta velocities, is fit with $m\approx 0.85$.  We
  infer a progenitor mass loss rate of $2\times 10^{-7}~\rm
  M_{\odot}~yr^{-1}$ (for wind velocity, $v_w=10^3~\rm km~s^{-1}$).
  These parameters constrain the characteristic synchrotron frequency,
  $\nu_m\approx 0.3$ GHz, and the synchrotron cooling frequency,
  $\nu_c\approx 10^{14}$ Hz, at $t=5$ days and thus $\nu_m < \nu_a$;
  consistent with the observed radio spectrum (inset, solid grey
  curve).  A nearly identical fit is obtained for a trans-relativistic
  GRB blastwave expanding into a constant density circumstellar
  medium\cite{gs02} for parameters: $E_K\approx 1.2\times 10^{48}$
  erg, $n=10^2~\rm cm^{-3}$, $\epsilon_e=\epsilon_B=0.1$ and $p=2.1$;
  in this case the mildly-relativistic ejecta is assumed to expand
  with a single bulk Lorentz factor.  These values constrain the
  geometry of the ejecta to be effectively spherical, $\theta_j\gtrsim
  1.4$. The X-ray flux (XRT=circles; XMM=encircled dot, scaled to XRT
  spectral model; CXO=squares) is significantly brighter than an
  extrapolation of the above model as evidenced by the unusually flat
  radio to X-ray spectral index, $\beta_{RX}\approx -0.5$ (inset,
  dashed line), and the steep X-ray spectrum $\beta_X \approx -2.2$
  (inset, black line), instead of $\beta_X\sim -1.1$ for typical GRBs.
  We suggest that the integrated optical to X-ray luminosity
  ($10^{44}~\rm erg~s^{-1}$; $2-10^4$ eV) can be attributed to the
  spin down power of a magnetar.  By day 5, the optical/UV spectrum is
  dominated by the thermal SN emission (inset).}
 
\bigskip

\noindent
{\bf Figure 2}: {\small Radio observations for a large sample of local
  Type Ibc supernovae.  Since 1999 we have been monitoring the radio
  emission from optically-selected SNe Ibc with the Very Large Array.
  We use radio luminosity as a proxy for mildly-relativistic ejecta to
  quantify the fraction of SNe Ibc powered by central engines.  Our
  observations of 144 SNe show that most SNe Ibc do not produce strong
  radio emission and therefore show no evidence for a central engine.
  For comparison, we include the radio afterglows for nearby
  ($z\lesssim 0.25$) GRBs 980425 and 030329, and XRF\,020903 all three
  of which show\cite{kfw+98,bkp+03,skb+04a} evidence for an
  engine-driven explosion.  XRF\,060218 is intermediate between GRBs
  and BL SN\,2002ap, demonstrating that broad lines are not a reliable
  proxy for strong radio emission.  Radio limits for other local
  broad-lined SNe (encircled triangles) show that less than one in
  three of these events may have a radio luminosity comparable to
  XRF\,060218 or GRB\,980425 ($90\%$ confidence level).  In addition,
  we show two 8.5 GHz model light-curves for a typical GRB viewed away
  from the collimation axis.  Both models adopt typical GRB parameters
  (see ref.~\pcite{snb+06} and references therein) of $\Gamma=100,
  E_{K,\rm iso}=10^{53}$ erg, $n=1~\rm cm^{-3}$,
  $\epsilon_e=\epsilon_B=0.1$, and $p=2.1$.  In the first model we
  assume that the observed $\gamma-$ray and radio emission are
  produced by a GRB jet viewed from an angle $\theta_{\rm
  obs}=2\theta_j$; here $\theta_{\rm obs}$ is the angle between our
  line-of-sight and the jet axis.  Under this framework, the observed
  prompt emission properties ($\Delta t$, $E_p$, $E_{\gamma,{\rm
  iso}}$) are related to the intrinsic values through the quantity
  $D\equiv[\Gamma(\theta_{\rm obs}-\theta_j)]^{-2}$.  For $D\sim
  0.02$, the intrinsic properties for XRF\,060218 would be typical for
  GRBs: $\Delta t\sim 40$ sec, $E_p\sim 250$ keV, $E_{\gamma,{\rm
  iso}}\sim 10^{53}$ erg, and $\theta_j\sim 4^{\circ}$.  The resulting
  off-axis model (dotted line) is a factor of $10^3$ brighter than the
  observed XRF\,060218 radio light-curve and can therefore be ruled
  out.  In the second model, we assume that in addition to the
  quasi-spherical mildly-relativistic ejecta component producing the
  observed radio emission, XRF\,060218 also harbors a strongly
  collimated relativistic jet directed significantly away from our
  line-of-sight.  In this scenario, we expect\cite{wax04,snb+06} a
  late-time radio re-brightening as the jet becomes non-relativistic
  and spreads sideways into our line-of-sight.  Adopting
  $\theta_j=4^{\circ}$ we find that our latest radio limit (104 days;
  black triangle) rules out an off-axis GRB with $\theta_{\rm
  obs}\lesssim 60^{\circ}$ (dash-dot line).  We conclude that the
  XRF\,060218 ejecta was quasi-spherical and intrinsically
  sub-energetic.}

\bigskip

\noindent
{\bf Figure 3}: {\small Energy as a function of velocity for GRBs,
  XRFs, and SNe Ibc.  Optical data (small dark circles) probe (see
  refs.~\pcite{bbh+99},\pcite{mdn+06} and references therein) the
  slowest ejecta in supernova explosions which typically carry the
  bulk of the kinetic energy ($E_K=0.3M_{\rm ej}v_{\rm ej}^2\sim
  10^{51}\,$erg).  On the other hand, radio observations (large light
  circles) trace\cite{bkp+03,kfw+98,skb+04b,bkf+03,c98} only the
  fastest ejecta in the explosion.  For GRBs 030329 and 031203,
  $\Gamma\propto t^{-3/8}$; we adopt the bulk velocity of the
  relativistic ejecta at day 1 as inferred from radio modeling.  For
  GRB\,980425, XRF\,060218, SN\,20020ap and SN\,1994I the bulk
  velocity is roughly constant on the timescale probed by the radio
  observations; we adopt the velocity at the radio peak time.
  Standard hydrodynamic collapse results\cite{tmm01} in a kinetic
  energy profile, $E_K\propto (\Gamma\beta)^{-5.2}$ (grey line), and
  thus a negligible fraction of the kinetic energy may be coupled to
  mildly-relativistic ejecta, consistent with the radio observations
  of local Ibc SNe 1994I and 2002ap.  In the case of typical GRBs
  (e.g. GRB\,030329), however, the kinetic energy of the
  mildly-relativistic ejecta is nearly comparable to that of the
  slower material indicating the presence of a central engine.  Since
  the origin of the relativistic flow is separate from the SN, there
  is probably not a continuous distribution of matter between the two
  data points but rather distinct ejecta components.  Sub-energetic
  bursts such as XRF\,060218 are intermediate between these two
  classes and may indicate that their central engines are different
  than those of typical GRBs.  We conclude that the minimum criteria
  for producing GRBs and XRFs is a mildly-relativistic ($\Gamma\gtrsim
  2$), quasi-spherical ejecta carrying at least $10^{48}$ erg.  }
\clearpage

\clearpage

\clearpage

\clearpage

\begin{center}
{\large \bf SUPPLEMENTARY INFORMATION}
\end{center}

\noindent
{\bf I. Estimates for the Rate of Sub-energetic GRBs}: To estimate the
rate of sub-energetic events similar to GRB\,980425 ($d=36.1$ Mpc) and
XRF\,060218 ($d=145$ Mpc), we consider only the satellite instruments
with precise localization capability: {\it BeppoSAX} Wide Field Cameras
(WFC), {\it High Energy Transient Explorer 2 (HETE-2)} Wide Field
X-ray Monitor (WXM), and {\it Swift} Burst Alert Telescope (BAT).
Inclusion of {\it INTEGRAL} would not significantly affect our results
due to its low GRB detection rate.  For these three instruments we
adopt the detection threshold curves calculated by Band (2003, ApJ,
588, 945; 2006, ApJ, 644, 378) in units of peak photon flux per second
($F_{\rm peak}$; 1-1000 keV) as a function of the $\nu F_{\nu}$
spectral peak energy, $E_{\rm peak}$.  We assume the shape of the
prompt emission spectrum is fit by a broken power-law (Band {\it et
al.} 1993, ApJ, 413, 281) such that $F_{\nu}\propto \nu^{\alpha}$
($F_{\nu}\propto \nu^{\beta}$) for $h\nu < E_{\rm peak}$ ($h\nu >
E_{\rm peak}$).  Using the observed spectral parameters ($F_{\rm
peak}$, $E_{\rm peak}$, $\alpha$, $\beta$) we calculate the
sensitivity of each instrument to each of the two events.

For GRB\,980425, $F_{\rm peak}\approx 3\times 10^{-7}~\rm
erg~cm^{-2}~s^{-1}$ (24-1820 keV; Galama {\it et al.} 1998, Nature,
395, 670) $E_{\rm peak}\approx 160$ keV, $\alpha\approx -0.27$ and
$\beta\approx -2$ (Jimenez, Band $\&$ Piran 2001, ApJ, 561, 171).
Extrapolating to the 1-1000 keV band, we find that the peak photon
flux is $F_p\approx 8.4~\rm ph~cm^{-2}~s^{-1}$ and thus the event
could be detected to 120 (WFC), 60 (WXM) and 130 (BAT) Mpc.

For XRF\,060218, $F_{\rm peak}\approx 5\times 10^{-8}~\rm
erg~cm^{-2}~s^{-1}$ (15-150 keV; Campana {\it et al.} 2006, Nature,
submitted) $E_{\rm peak}\approx 4.9$ keV, $\beta\approx -1.5$ and
$\alpha$ is not constrained (15-150 keV; Campana {\it et al.} 2006,
Nature, submitted).  We estimate $F_p\approx 1.9~\rm
ph~cm^{-2}~s^{-1}$ and therefore XRF\,060218 could be detected to 220
(WFC), 110 (WXM) and 180 (BAT) Mpc.  We note that unlike other
missions, the BAT detection threshold is significantly lower (factor
of $\sim 3$) for unusually long duration events such as XRF\,060218.

Next we estimate the effective monitoring time, $T_m$ of each of the
missions assuming their sky coverage, $S$ and operation time, $T$.
For the two Wide Field Cameras $S=2\times 0.123=0.246$~sr (Band, 2003,
ApJ, 588, 945) and $T=4$ yrs, and for WXM $S=0.806$~sr and $T=3$ yrs
(Guetta {\it et al.}, 2004, ApJ, 615, L73).  For BAT, $S=2$~sr and
$T=1$ yr (S. Barthelmy, private communication).  Thus we find
monitoring times, $T_m=(T/4\pi)S$, of 0.08 (WFC), 0.19 (WXM), and 0.16
(BAT) yrs.

We estimate the sensitivity of these instruments to each of the events
as $T_m\times V$; here $V$ is the volume to which each event could be
detected.  Adopting the larger of the two sensitivities for each
instrument we find $3.8\times 10^{-3}$ (WFC), $1.2\times 10^{-3}$
(WXM) and $3.7\times 10^{-3}$ (BAT) Gpc$^3$ yr.  Summing the
sensitivities, we find that the rate of sub-energetic events is
$230^{+490}_{-190}~~\rm Gpc^{-3}~yr^{-1}$ where the errors are
dominated by the $90\%$ Poisson statistics for two detections
(Gehrels, 1986, ApJ, 303, 336).

\noindent
{\bf II. Estimates for the rate of Type Ibc supernovae like GRB\,980425
and XRF\,060218}: To estimate the rate of SNe Ibc with strong, early
radio emission comparable to that observed for sub-energetic bursts we
only consider the 75 events (out of 144 optically-selected local SNe
Ibc) with $3\sigma$ upper limits fainter than the observed GRB\,980425
and XRF\,060218 light-curves at that same epoch.  We then assume
various values for the true fraction of SNe Ibc with radio emission
comparable (or higher) to that of XRF\,060218 and GRB\,980425 and
determine the probability of finding null-detections for all 75 events
for each assumed fraction.  Larger fractions are ruled out with higher
confidence.  At 90\% confidence, we rule out the scenario where
$\gtrsim 3\%$ of SNe Ibc are as radio bright as XRF\,060218 and
GRB\,980425.  Adopting the local rate of SNe Ibc, $9^{+3}_{-5}\times
10^{-3}~\rm Gpc^{-3}~yr^{-1}$, as measured by Cappellaro {\it et al.}
(1999, Astr. Astrophysics, 351, 459) and Dahlen {\it et al.} (2004,
ApJ, 613, 189), we conclude that the volumetric rate of events like
GRB\,980425 and XRF\,060218 is less than $3\%$ of the local SNe Ibc
sample, or $\lesssim 300~\rm Gpc^{-3}~yr^{-1}$.

Repeating this analysis for the subset of broad-lined SNe Ibc, we
find that at $90\%$ confidence we can rule out the scenario where $\gtrsim
30\%$ of local, optically selected BL SNe Ibc produce radio emission
similar to that observed for GRB\,980425 and XRF\,060218.

\end{document}